\documentclass[12pt,preprint]{aastex}

\def\lesssim{\mathrel{\hbox{\rlap{\hbox{\lower4pt\hbox{$\sim$}}}\hbox{$<$}}}}
\def\gtrsim{\mathrel{\hbox{\rlap{\hbox{\lower4pt\hbox{$\sim$}}}\hbox{$>$}}}}
\providecommand{\etal}{et~al.}

\begin{document}
\title{GRB 060313: A New Paradigm for Short-Hard Bursts?}
\author{Peter W.~A. Roming$^1$, Daniel Vanden Berk$^1$, Valentin 
Pal'shin$^2$, Claudio Pagani$^1$, Jay Norris$^3$, Pawan Kumar$^4$,
Hans Krimm$^{3,5}$, Stephen T. Holland$^{3,5}$, Caryl Gronwall$^1$, 
Alex J. Blustin$^6$, Bing Zhang$^7$, Patricia Schady$^{1,6}$, 
Takanori Sakamoto$^3$, Julian P. Osborne$^8$, John A. Nousek$^1$, 
Frank E. Marshall$^3$, 
Peter M\'{e}sz\'{a}ros$^{1,9}$, Sergey V. Golenetskii$^2$, Neil 
Gehrels$^3$, Dmitry D. Frederiks$^2$, Sergio Campana$^{10}$, David N. 
Burrows$^1$, Patricia T. Boyd$^3$, Scott Barthelmy$^3$, 
R.~L. Aptekar$^2$} 
\affil{$^1$ Department of Astronomy \& Astrophysics,
Pennsylvania State University, 525 Davey Lab, University Park, PA
16802, USA; Corresponding author's e-mail: roming@astro.psu.edu\\
$^2$ Ioffe Physico-Technical Institute, 26 Polytekhnicheskaya, St. 
Petersburg 194021, Russian Federation\\ 
$^3$ NASA/Goddard Space Flight Center, Greenbelt, MD 20771, USA\\
$^4$ Department of Astronomy, University of Texas at Austin, 1 
University Station, C1400 Austin, Texas 78712-0259, USA\\
$^5$ Universities Space Research Association, 10227 Wincopin Circle, 
Suite 500, Columbia, MD 21044, USA\\ 
$^6$ Mullard Space Science Laboratory, University College London, 
Holmbury St. Mary, Dorking, Surrey RH5 6NT, UK\\ 
$^7$ Department of Physics, University of Nevada, Las Vegas, 4505 
Maryland Parkway, Las Vegas, NV 89154-4002, USA\\ 
$^8$ Department of Physics and Astronomy, University of Leicester, 
University Road, Leicester LE1 7RH, UK\\
$^9$ Department of Physics, Pennsylvania State University, 104 Davey 
Lab, University Park, PA 16802, USA\\ 
$^{10}$ INAF ­ Osservatorio Astronomico di Brera, via E. Bianchi 46, 
I­23807 Merate (LC), Italy}

\begin{abstract}
We report the simultaneous observations of the prompt emission in the 
$\gamma$-ray and hard X-ray bands by the Swift-BAT and the KONUS-Wind 
instruments of the short-hard burst, GRB 060313. The observations 
reveal multiple peaks in both the $\gamma$-ray and hard X-ray bands 
suggesting 
a highly variable outflow from the central explosion. We also describe 
the early-time observations of the X-ray and UV/Optical afterglows by 
the Swift XRT and UVOT instruments. The combination of the X-ray and 
UV/Optical observations provide the most comprehensive lightcurves to 
date of a short-hard burst at such an early epoch. The afterglows 
exhibit complex structure with different decay indices and 
flaring. This behavior can be explained by the combination of a structured 
jet, radiative loss of energy, and decreasing microphysics parameters 
occurring in a circum-burst medium with densities varying by a factor of 
approximately two on a length scale of $10^{17} {\rm ~cm}$. 
These density variations are normally associated with the environment 
of a massive star and inhomogeneities in its windy medium. However, the 
mean density of the observed medium ($n \sim 10^{-4} {\rm ~cm^{3}}$) is 
much less than that expected for a massive star. Although the collapse 
of a massive star as the origin of GRB 060313 is unlikely, the merger of 
a compact binary also poses problems for explaining the behavior of this 
burst. Two possible suggestions for explaining this scenario are: some 
short bursts may arise from a mechanism that does not invoke the 
conventional compact binary model, or soft late-time central engine 
activity is producing UV/optical but no X-ray flaring.
\end{abstract}

\keywords{gamma-rays: bursts}

\section {Introduction}
Gamma-ray bursts (GRBs) are generally classified into one of two 
categories: long-soft and short-hard \citep{KC1993}\footnote{Two 
additional classes of bursts have been proposed: an intermediate class
\citep{HI1998, MS1998, HI2006} which falls between the short and long
burst classes with respect to $T_{90}$ and has a hardness ratio softer than 
long bursts, and a very short class \citep{CDB2005} which has a very hard 
spectrum and $T_{90} \leq 100 {\rm ~ms}$.}. Short GRBs are 
bursts with $T_{90} < 2 {\rm ~s}$ (where $T_{90}$ is the time in which the 
cumulative counts above background are between 5\% and 95\% of the total) 
and a hardness ratio that is typically larger than the one of long GRBs. 
Short bursts 
also exhibit an initial spike with insignificant spectral evolution at 
energies greater than $\sim 25 {\rm ~keV}$ \citep{NB2006}. 25-30\% of all 
GRBs in the Burst and Transient Source Experiment 
(BATSE) catalog \citep{FG1994} are short-hard bursts \citep[SHBs;][]{KC1993}, 
while 12$\pm$4\% of the Swift \citep{GN2004} sample are SHBs. It has been 
proposed that SHBs are the result of a compact binary merger such as a 
double neutron star \citep[DNS;][]{PB1986, ED1989, NPP1992, FWH1999,
RRD2003, LRG2005, PA2006a} or a 
neutron star and a black hole \citep[NS-BH;][]{LS1976, PB1991, 
NPP1992, MR1993, FWH1999, LRG2005, PA2006a}. Since the natal kick of a 
neutron star is $\sim 200-1000 
{\rm ~km \,s^{-1}}$, and since the typical life time of compact binaries is 
$\sim 0.1-1 {\rm ~Gyr}$ \citep{PA2006a}, a compact binary could wander anywhere 
from $20 {\rm ~kpc}$ to $1 {\rm ~Mpc}$ in the time it takes for the binary 
to merge. At these 
large distances from the neutron star's stellar nursery, a lower density 
external medium is expected.

Understanding and testing these different short GRB models has been 
difficult due to the large time delay in localizing the burst afterglows. 
The Interplanetary Network \citep[IPN;][]{HK2005} has triangulated a few 
short GRBs on timescales of days \citep{BJ2006}; however, on these 
timescales, the afterglow has long since faded. Since the launch of the 
Swift satellite, eleven SHBs\footnote{It has been noted in the 
literature that there have been other 
well localized short bursts which have soft spectra \citep{BJ2006}. Recent 
work \citep{LA2006, ST2006} has suggested that 
Swift and HETE-2 short bursts have softer spectra than the BATSE bursts. 
They find that the hardness ratio is in-between that of the BATSE short and 
long GRBs, but the Swift short bursts are typically harder than the long 
ones. Hereafter, we assume the Swift and HETE-2 short bursts fit into the 
hard classification scheme and refer to short bursts as those bursts with 
short-hard spectra.} 
have been localized in under an hour. Four 
of the eleven bursts, GRBs 050202 \citep{TJ2005}, 050906 \citep{KH2005}, 
050925 \citep{HST2005}, and 051105A \citep{MT2005}, were localized in 
the hard X-ray band by the Swift Burst Alert Telescope \citep[BAT;][]{BS2005a} 
but with no corresponding X-ray, optical, or radio detections; three 
additional bursts,  GRBs 050509B \citep{GN2005, BJ2006}, 050813 
\citep{RA2005, MD2005}, and 051210 \citep{LPV2006}, were 
localized by the BAT with corresponding Swift X-ray Telescope 
\citep[XRT;][]{BDN2005a} detections but no optical or radio detections; the 
remaining four bursts, GRBs 050724 \citep{BS2005b, BE2005, CS2006, GD2006, 
VS2006}, 051221A \citep{PA2005, BCG2005, GBP2005, RPWA2005a, SA2006, 
BDN2006}, 051227 \citep{BL2006}, and 060313 \citep{PC2006a}, 
were localized by the BAT and XRT with corresponding Swift 
Ultra-Violet/Optical Telescope \citep[UVOT;][]{RPWA2005b} and/or ground-based 
detections. Additional rapid observations of GRBs 050709 \citep{FD2005, VJ2005, 
CSEA2006} and 060121 \citep{AM2006, LA2006} 
were made with the High Energy Transient Explorer (HETE-2). These rapid 
observations have broadened our understanding of short GRBs and provided 
strong evidence that the origin of short bursts is different from that of 
long bursts and is consistent with the merger 
of compact objects \citep{GN2005, BS2005b, FD2005, VJ2005}.

However, the data for GRB 060313 indicates that there are problems with the model 
for a compact binary merger, at least for this burst. Two possible explanatons 
for the observed behavior are: some short bursts may be produced by a different 
mechanism than that invoked by the standard compact binary model, or late-time 
central engine activity is generating late internal shocks from which low energy 
(UV/optical) flares arise. Here we 
present the broad-band nature of the short GRB 060313 and its afterglow based 
on Swift BAT, XRT, and UVOT as well as KONUS-Wind \citep{AR1995} data. This 
is only the fifth short burst with a reported optical afterglow and the first 
with UV detections and optical flaring. It is also the hardest burst in the 
Swift sample. In addition, the optical and UV detections made by the UVOT of 
this burst are the earliest ($< 80 {\rm ~s}$) optical/UV detections of a short burst to 
date. The combination of the UVOT data with the BAT, KONUS, and XRT data provide 
the most comprehensive lightcurves to date of a SHB at such an early epoch and 
affords a unique opportunity to probe the mechanism behind this short GRB. 

\section{Observations \& Data Analysis}
On March 13, 2006, at $00:12:06.484 {\rm ~UT}$, the BAT triggered on GRB 060313 
\citep{PC2006a}. The timing analysis hereafter is referenced from the BAT 
trigger time ($T_0$). The initial BAT light curve has two slightly 
overlapping peaks with a total duration of $\sim 1 {\rm ~s}$. The peak count rate was 
approximately $90,000 {\rm ~counts\,s^{-1}}$ ($15-350 {\rm ~keV}$) occurring at $T_0+0.5$\,s. 

The UVOT began a finding chart exposure $78 {\rm ~s}$ after the trigger. The afterglow 
was discovered during ground processing. The XRT began its autonomous sequence 
of observations of the GRB field at $00:13:24 {\rm ~UT}$, $79 {\rm ~s}$ after the BAT trigger. 
The XRT on-board centroiding algorithm could not converge on a source in the 
image, but a fading source was detected in the ground-processed data 
\citep{PC2006b}. The first ground based detection was made at $01:28 {\rm ~UT}$ with 
the VLT and FORS2 telescopes \citep{LH2006}. No radio source was detected at 
the VLT position \citep{SF2006}.

Hereafter, we adopt the notation $F(\nu ,t) \propto t^{-\alpha} \nu^{-\beta}$ 
for the afterglow flux as a function of time, where $\nu$ is the frequency of 
the observed flux, $t$ is the time post trigger, $\beta$ is the spectral index 
which is related to the photon index $\Gamma$ ($\beta = \Gamma - 1$) , and 
$\alpha$ is the temporal decay slope. We also adopt the convention $q_x = 10^x q$ in cgs units.
A flat cosmological constant dominated cosmology with the parameter values 
$H_0 = 70 {\rm ~km\,s^{-1}\,Mpc^{-1}}$, $\Omega_M = 0.3$, and $\Omega_{\Lambda} = 0.7$
is assumed.

\subsection{BAT Analysis}
Ground analysis \citep{MC2006} of the BAT data determined that $T_{90}$ for 
the $15-350 {\rm ~keV}$ band is $0.7\pm0.1 {\rm ~s}$ (estimated error includes the systematics) 
with a fluence of $1.13\pm0.05 \times 10^{-6} {\rm ~erg\,cm^{-2}}$ in the $15-150 {\rm ~keV}$ 
band. The 1 second peak photon flux measured from $T_0-0.124$ s in the $15-150 {\rm
~keV}$ band is $12.1\pm0.4 {\rm ~ph\,cm^{-2}\,s^{-1}}$. All BAT errors are at the 90\% 
confidence level.

\begin{figure}
\plotone{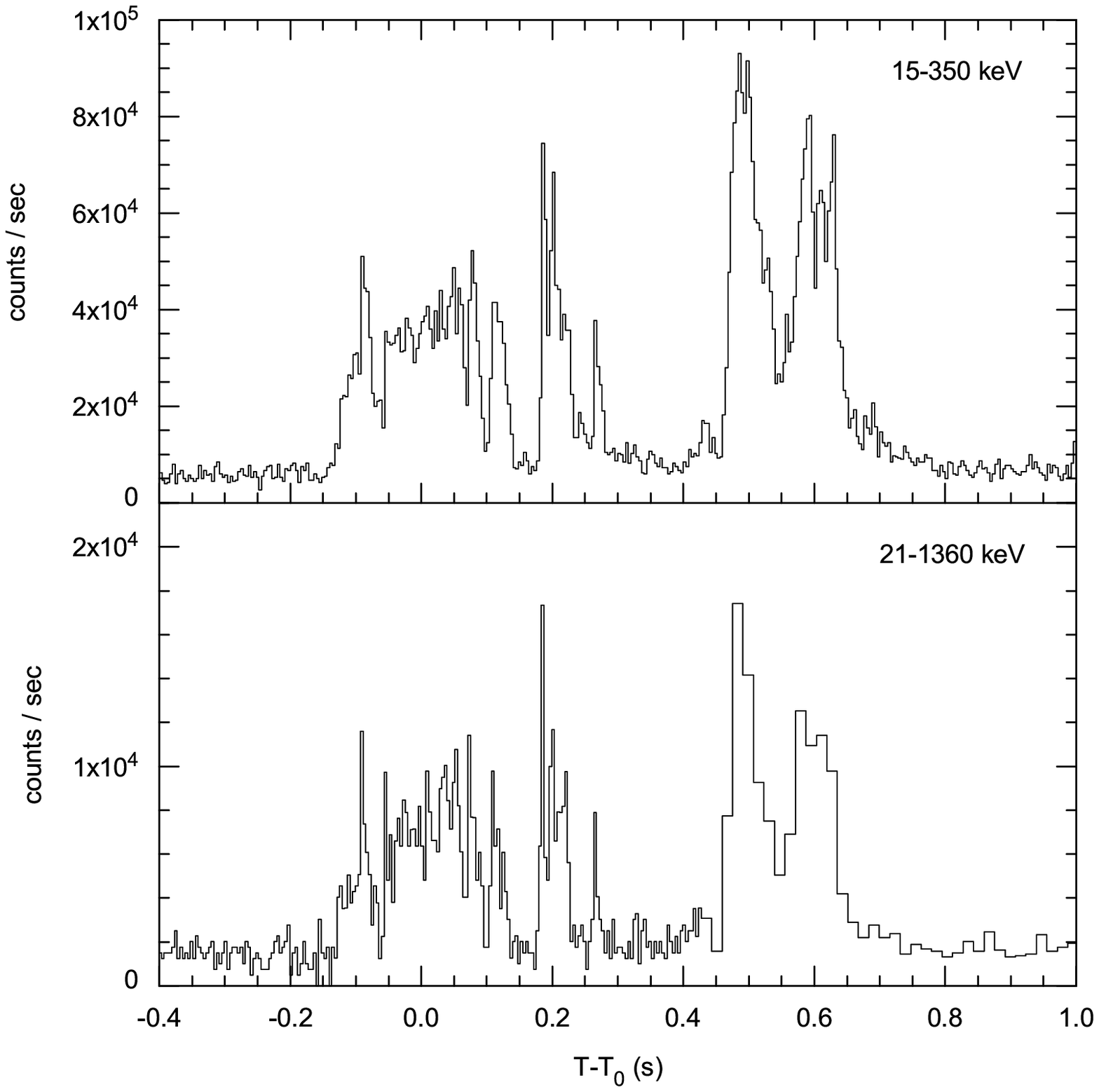} 
\caption{BAT and KONUS-Wind lightcurves. The BAT $15-350 {\rm ~keV}$ lightcurve 
(upper panel) has at least 20 statistically significant peaks with FWHMs 
in the $5-15 {\rm ~ms}$ range. The average BAT error is $1250 {\rm counts s^{-1}}$. The 
KONUS-Wind $21-1360 {\rm ~keV}$ lightcurve (lower panel) also exhibits several 
multi-peaked pulses with a total duration of approximately $0.8 {\rm ~s}$.\label{fig1}}
\end{figure}

In the $15-350 {\rm ~keV}$ lightcurve (Figure~\ref{fig1}) at least 20 statistically significant 
peaks with FWHMs in the $5-15 {\rm ~ms}$ range \citep{BS2006} can be seen. There is 
no periodic structure in the lightcurve for at least the first $100 {\rm ~s}$.  There 
is also no evidence of extended emission in the $T_0+1$ to $T_0+300 {\rm ~s}$ range 
at an upper limit of $0.001 {\rm ~counts ~detector^{-1} ~s^{-1}}$ ($3\sigma$).  This is consistent 
with an upper limit on the flux ratio between the initial peak and the peak 
of any potential extended emission of $\sim 2000$. SHBs 050724 and 051227 had flux 
ratios of 46 and $\sim 10$, respectively \citep{BS2006}.

\begin{figure}
\epsscale{0.8}
\plotone{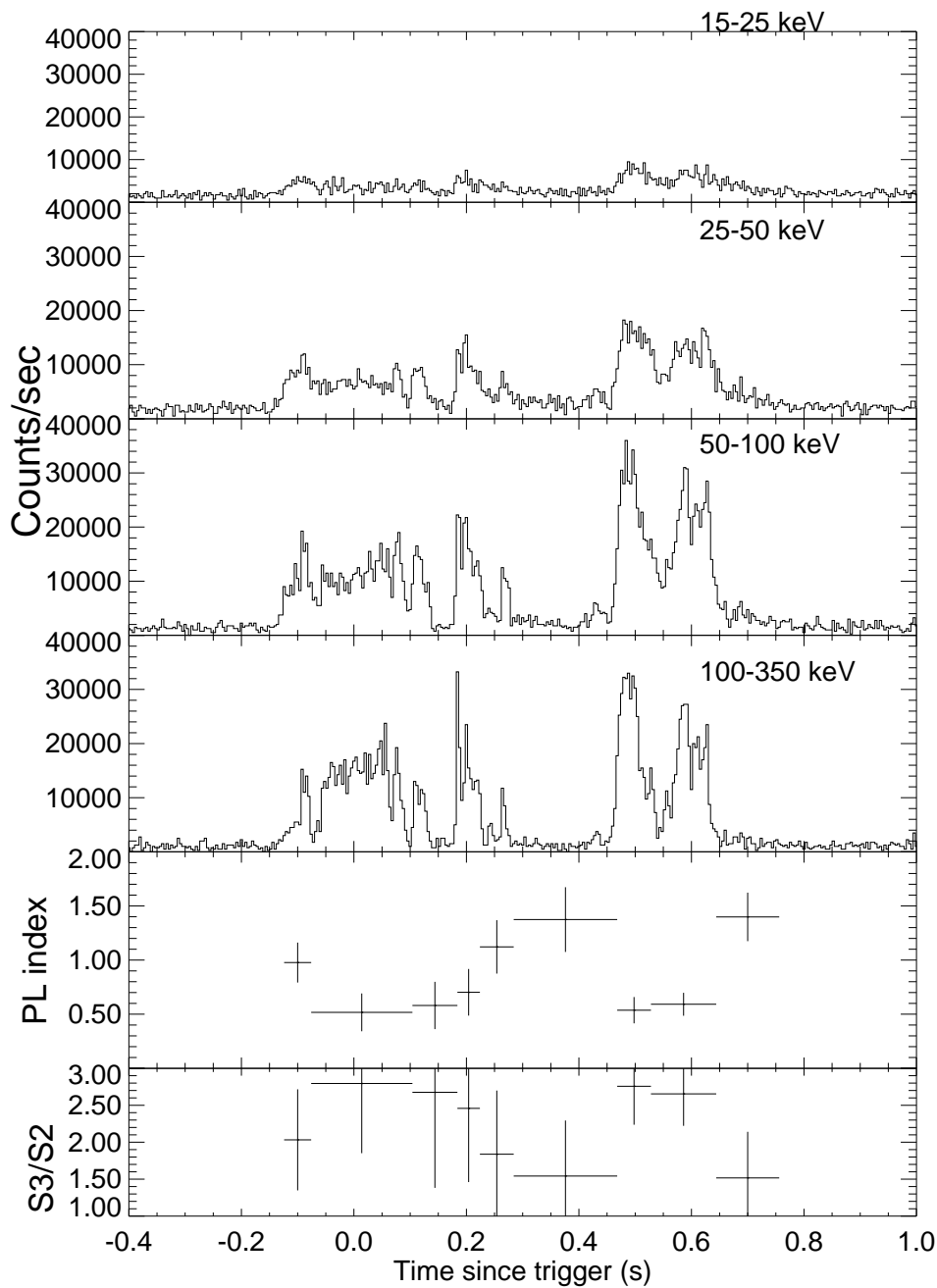} 
\caption{BAT four channel ($15-25$, $25-50$, $50-100$, \& $100-350 {\rm ~keV}$) light 
curves. The average errors are 655, 670, 600, and $530 {\rm ~counts s^{-1}}$, 
respectively. The fifth panel from the top shows the photon power-law 
index for a simple power law fit. The bottom panel shows the fluence 
hardness ratio ${\rm S3/S2}$ between the $50-100 {\rm ~keV}$ and $25-50 {\rm ~keV}$ 
bands, where 
the fluence was calculated based on a power-law fit. The hardness ratio 
for the time integrated spectral fit is 2.43.\label{fig2}}
\end{figure}

A four channel lightcurve (Figure~\ref{fig2}) reveals that the burst is a short hard 
burst. A lag analysis confirmed the burst to cleanly reside in the short hard 
burst class \citep[Figure 3 in][]{NB2006}. The measured lags are $0.8\pm0.6 {\rm ~ms}$ 
$[(50-100 {\rm ~keV})/(15-25 {\rm ~keV})]$ and $0.3\pm0.7 {\rm ~ms}$ 
$[(100-350 {\rm ~keV})/(25-50 {\rm ~keV})]$.

\subsection{KONUS-Wind Analysis}
KONUS-Wind (K-W) triggered on GRB 060313 at $T_0(K-W) = 00:12:06.354 {\rm ~UT}$ 
\citep{GS2006}. It was detected by the S1 detector which observes the south ecliptic 
hemisphere; the incident angle was $57.9\arcdeg$. The propagation delay from 
Swift to Wind is $0.045 {\rm ~s}$ for this GRB, therefore, correcting for this factor, 
one sees that the K-W trigger time corresponds to $T_0-0.085 {\rm ~s}$.

\begin{figure}
\epsscale{0.9}
\plotone{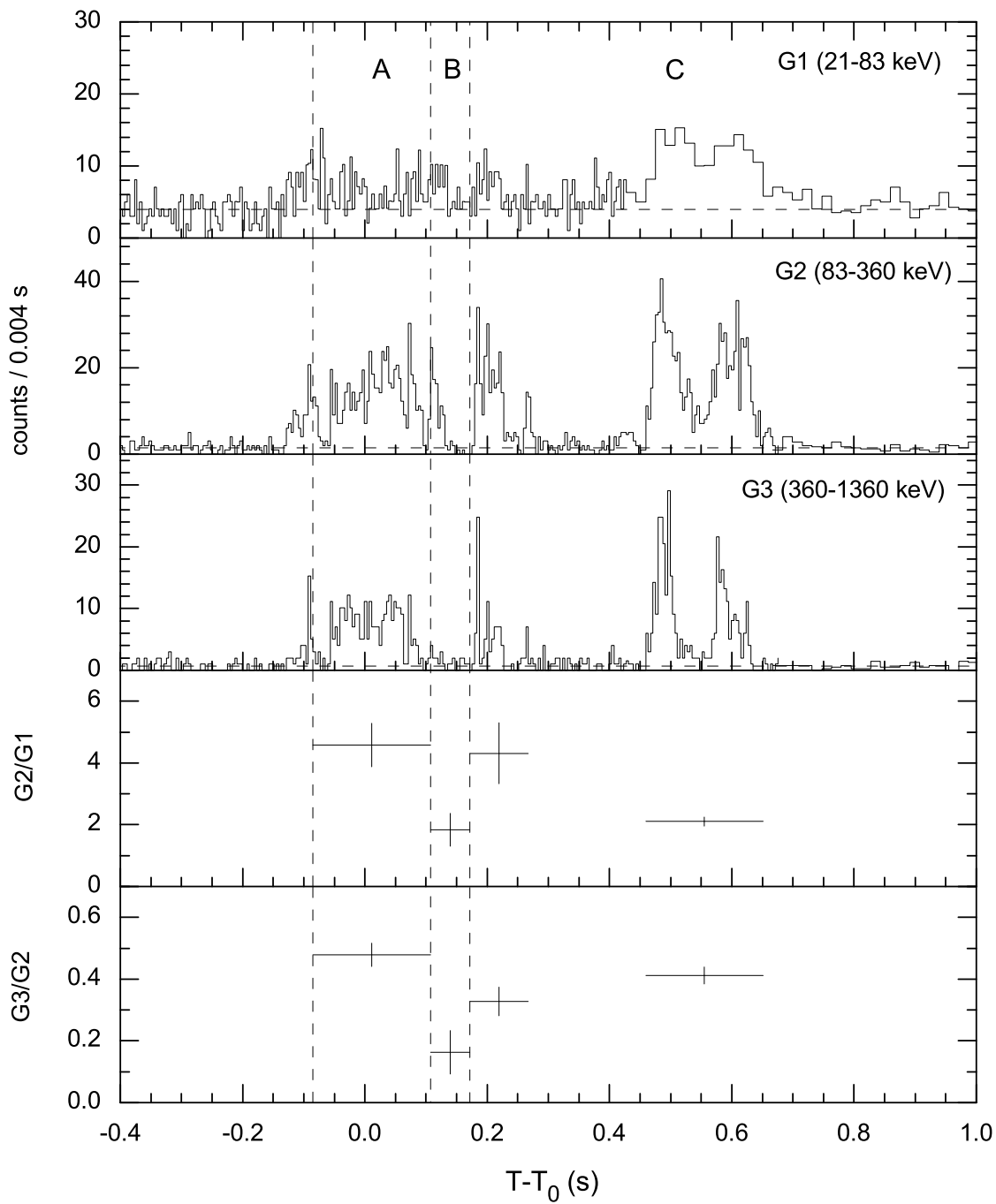} 
\caption{KONUS-Wind lightcurve of GRB 060313 in three energy bands. 
The two bottom panels show the hardness ratios ${\rm G2/G1}$ and ${\rm G3/G2}$ which 
demonstrates significant spectral evolution during the burst. The 
dashed vertical lines denote the boundaries of the three intervals 
over which we extracted the spectra in Section 2.4.\label{fig3}}
\end{figure}

The GRB time history was recorded in three energy ranges: G1 ($21-83 {\rm ~keV}$), G2 
($83-360 {\rm ~keV}$), and G3 ($360-1360 {\rm ~keV}$). Thanks to the high intensity of the burst, 
the K-W reserve system (so called 'nonius') was triggered providing $2 {\rm ~ms}$ 
resolution record in the G2 and G3 ranges of the entire burst (up to 
$T_0(K-W)+0.762 {\rm ~s}$). Five spectra in 101 channels were accumulated during the 
burst (the total number of spectra accumulated during the trigger record is 64). 
The light curve (Figure~\ref{fig1}) shows several multi-peaked pulses with a total duration 
of approximately $0.8 {\rm ~s}$. As with the BAT multi-channel light curve, the three 
channel KONUS light curves (Figure~\ref{fig3}) illustrate that the burst is a short 
hard burst.

\subsection{XRT Analysis}
\begin{figure}
\plotone{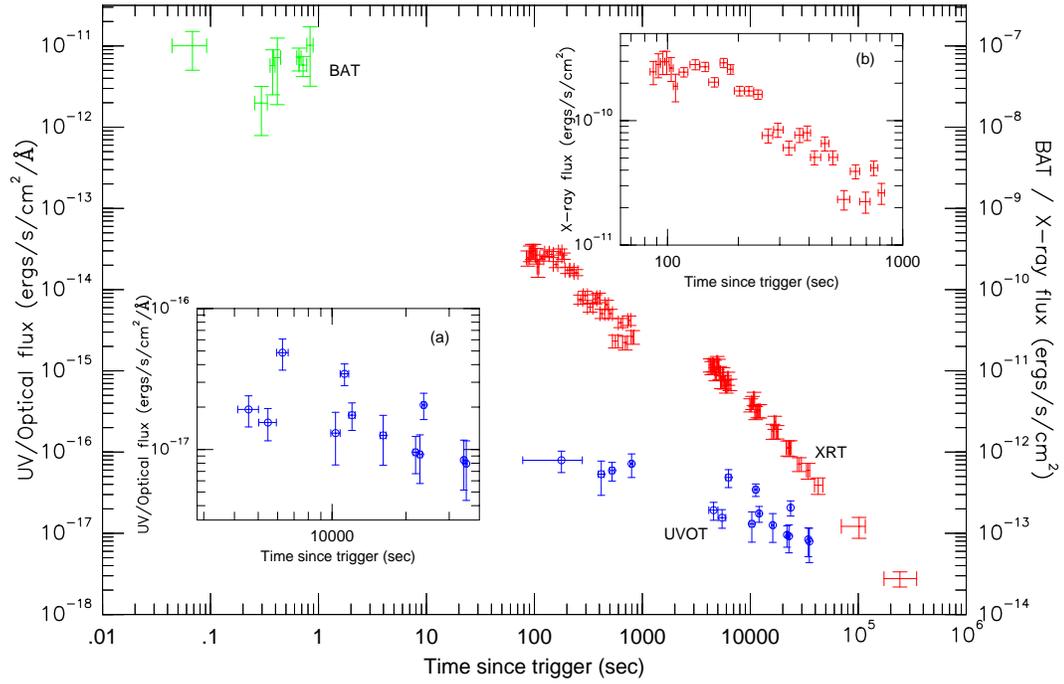} 
\caption{Combined BAT, XRT, and UVOT lightcurves. The green represents 
the BAT values extrapolated into the XRT energy range, the red 
represents the XRT $0.3-10 {\rm ~keV}$ fluxes, and the blue represent the UVOT 
values adjusted to the $U$-band. The two insets are provided to show 
the flaring of the UVOT (insert (a)) and X-ray (insert (b)) lightcurves 
after $3000 {\rm ~s}$ and during the first $1000 {\rm ~s}$, respectively.\label{fig4}}
\end{figure}

An uncatalogued, fading source was discovered during ground analysis \citep{PC2006b} 
of the XRT data. The X-ray afterglow was observed in Windowed Timing (WT) mode for 
$26 {\rm ~s}$. As the source faded the XRT switched into Photon Counting (PC) mode, thus 
providing 2-dimensional spatial information \citep[for a complete description of the
XRT modes see][]{HJ2004}. To avoid pile-up in the PC data during the first orbit 
of observations, the XRT lightcurve was extracted in the energy range $0.2-10.0 {\rm ~keV}$ 
using an annulus centered on the afterglow position with an inner radius of 2 pixels 
and applying a PSF correction. The X-ray lightcurve (see Figure~\ref{fig4}) during the first 
orbit of observation manifested variability, with spikes and fluctuations superimposed 
on the overall fading behavior. The initial decay slope was $\alpha = 1.25\pm0.15$.
The lightcurve profile becomes smoother in the later 
orbits ($> 4200 {\rm ~s}$ after the BAT trigger) and the decay curve can be fitted 
by a power law with index of $1.46\pm0.08$. In addition, the flux at these later 
times is larger by a factor $\sim 4$ compared with the extrapolation of the earlier 
X-ray lightcurve. The X-ray lightcurve is shown in Figure~\ref{fig4}, with the 
count rate converted into $0.2-10.0 {\rm ~keV}$ unabsorbed flux using the best 
spectral model fit.

The X-ray spectrum from the Photon Counting data was extracted in the $0.3-10.0 {\rm ~keV}$ 
energy range, with a binning of at least 20 counts per energy bin, using the latest 
versions of the ancillary and response matrix files. The PC spectrum of the first 
orbit of observation was extracted excluding the central 2 pixels to avoid pile up. 
The spectrum can be fitted by a simple absorbed power law, yielding a photon index 
of $1.53\pm0.10$ and an absorption column density $N_H$ consistent with the Galactic 
value of $4.7\times 10^{20} {\rm ~cm^{-2}}$ \citep{DL1990}. The spectrum softens during 
the later orbits: the best power law model fit of the afterglow spectrum extracted 
from 4.1 to $24 {\rm ~ks}$ after the GRB trigger yields a photon index of $1.96\pm0.09$, 
fixing the $N_H$ at the Galactic value.

\subsection{UVOT Analysis}
The Swift spacecraft slewed promptly when the BAT detected GRB 060313, 
and UVOT began imaging the field shortly after the BAT trigger. The 
UVOT took 44 exposures of the field between 78 and $67\,783 {\rm ~s}$ after 
the BAT trigger. For the majority of these exposures the afterglow was 
too faint to be detected. The UVOT photometry that is used in this 
paper is presented in Table~\ref{tab1}. The UVW2 detection (see below) 
implies that the 
Lyman limit must be blueward of $\sim 250 {\rm ~nm}$, suggesting an upper 
limit on the redshift of $z \lesssim 1.7$. To further constrain the 
redshift, $z$, a power-law UV/optical SED with an unconstrained 
spectral slope, $\beta$, as well as a fixed Galactic E($B-V$) was 
assumed. The redshift was then varied and the spectrum for a given 
$\beta$ was modified according to the parameterization of the average 
Lyman absorption \citep{MP1995}. The magnitudes relative to each UVOT 
filter were then estimated for a grid of $\beta$ and $z$, with a 
corresponding $\chi^2$ value for each $\beta$-$z$ pair. The grid of 
$\chi^2$ values were plotted to illustrate how $\beta$ and $z$ could 
be constrained. The probability distributions were also projected onto 
each axis to obtain the confidence intervals. Neither parameter is very 
well constrained, but the best $\beta \sim 1.93\pm0.22$ and $z < 1.1$ at 
the 90\% confidence level. The most likely redshift is $z = 0.75$, but 
it is minimally peaked.

\begin{deluxetable}{rrclll}
\tablecolumns{6}
\tabletypesize{\small}
\tablecaption{{\em Swift}/UVOT photometry of the afterglow of
	GRB 060313\label{tab1}}
\tablewidth{0pt}
\tablehead{
  \colhead{$t$ (s)} &
  \colhead{$\Delta t$ (s)} &
  \colhead{Filter} &
  \colhead{Mag} &
  \colhead{Error} &
  \colhead{Adjusted $U$-band Mag}
}
\startdata
178 & 200 & $V$ & 19.56 & 0.32 & 19.06 \\
416 & 50 & $U$ & 19.49 & 0.49 & 19.49 \\
524 & 50 & White & 19.07 & 0.28 & 19.38 \\
795 & 50 & $U$ & 19.17 & 0.35 & 19.17 \\
4568 & 886 & $B$ & 20.85 & 0.27 & 20.60 \\
5476 & 900 & UVW2 & 20.20 & 0.28 & 20.83 \\
6280 & 684 & $V$ & 20.09 & 0.27 & 19.59 \\
10328 & 900 & UVM2 & 20.53 & 0.44 & 21.02 \\
11235 & 900 & UVW1 & 19.66 & 0.19 & 19.97 \\
12052 & 708 & $U$ & 20.70 & 0.24 & 20.70 \\
16147 & 886 & $B$ & 21.31 & 0.42 & 21.06 \\
21890 & 900 & UVM2 & 20.87 & 0.32 & 21.36 \\
22797 & 900 & UVW1 & 21.09 & 0.41 & 21.40 \\
23618 & 716 & $U$ & 20.52 & 0.23 & 20.52 \\
34426 & 900 & UVW1 & 21.19 & 0.42 & 21.50 \\
35217 & 658 & $U$ & 21.56 & 0.49 & 21.56 \\
\enddata
\end{deluxetable}

\begin{deluxetable}{cll}
\tablecolumns{3}
\tabletypesize{\small}
\tablecaption{Zero points applied to the photometry of the afterglow 
	of GRB~060313\label{tab2}}
\tablewidth{0pt}
\tablehead{
  \colhead{Filter} &
  \colhead{ZP} &
  \colhead{Error}
}
\startdata
$V$ & 17.88 & 0.09 \\
$B$ & 19.16 & 0.12 \\
$U$ & 18.38 & 0.23 \\
UVW1 & 17.69 & 0.02 \\
UVM2 & 17.29 & 0.23 \\
UVW2 & 17.77 & 0.02 \\
White & 19.78 & 0.02 \\
\enddata
\end{deluxetable}

The UVOT has $UBV$ filters which approximate the Johnson system and 
three ultraviolet filters: UVW1 with a central wavelength of 
$\lambda_c = 251 {\rm ~nm}$, UVM2 with $\lambda_c = 217 {\rm ~nm}$, 
and UVW2 with 
$\lambda_c = 188 {\rm ~nm}$. Examination of the first settled observation 
(a $200 {\rm ~s}$ exposure in $V$) revealed a new source relative to the 
Digital Sky Survey inside the XRT error circle. This source had a 
magnitude of $V = 19.56\pm0.32$. The UVOT position of the source is 
$RA = 04h\,26m\,28.429s$ and $Dec = -10\arcdeg\,50\arcmin\,39.13\arcsec$ 
(J2000), with an internal accuracy of $\pm0\farcs01$ and an 
absolute astrometric uncertainty of $0\farcs56$ (90\% containment). 
This position is $1\farcs3$ from the reported XRT position 
\citep{PC2006b}. It is inside the XRT error circle and consistent 
with reported ground-based detections \citep{LH2006,TC2006}. 
Subsequent exposures showed the source to be 
fading.

We performed photometry on each UVOT exposure using a circular 
aperture with a radius of 2\arcsec ~centered on the position of the 
optical afterglow. This radius is approximately equal to the FWHM of 
the UVOT PSF. The PSF varies with filter and with spacecraft 
voltage, so we did not match the extraction aperture to the PSF for 
each exposure. The PSF FWHM, averaged over the voltage variations, 
ranges from $1\farcs79\pm0\farcs05$ for the $V$ filter to 
$2\farcs17\pm0\farcs03$ for the UVW2 filter. The background was 
measured in a sky annulus of inner radius $17\farcs5$ and width 
5\arcsec ~centered on the afterglow.

Aperture corrections were computed for each exposure to convert the 
2\arcsec ~photometry to the standard aperture radii used to define 
UVOT's photometric zero points (6\arcsec ~for $UBV$ and 12\arcsec ~for 
the UV filters). The aperture correction procedure also accommodates 
the variable PSF. Approximately six isolated stars were used to 
compute the aperture correction for each exposure. The RMS scatter 
in the mean aperture correction for a single exposure was typically 
$\sim 0.07$ mag. The RMS scatter for each exposure was added in 
quadrature to the statistical error in the 2\arcsec ~magnitude to 
obtain the total $1\sigma$ error in each point. All detections above 
the $2\sigma$ significance level are tabulated in Table~\ref{tab1}.

Since the UVOT is a photon-counting device, we have corrected all of 
our data for coincidence loss; however, the afterglow has $V > 19$, 
so coincidence losses are negligible, typically less than 0.01 mag. 
The zero points used to transform the instrumental UVOT magnitudes 
to Vega magnitudes are listed in Table~\ref{tab2}. They are taken from the 
latest in-orbit measurements as obtained from the HEASARC Swift/UVOT 
Calibration Database 
(CalDB)\footnote{http://swift.gsfc.nasa.gov/docs/heasarc/caldb/swift/}. 
Color terms were not applied to the photometric calibrations, but 
preliminary calibrations of on-orbit data suggest that they are 
negligible.

The UVOT photometry was adjusted to the $U$-band by assuming that 
the optical spectrum was a power law with the same slope as the 
X-ray spectrum ($\beta_X = 0.96$). The resultant UVOT photometry is 
provided in Table~\ref{tab1}, Adjusted $U$-band Mag column, and is shown 
in Figure~\ref{fig4}. In addition, the UVOT photometry was also adjusted to 
the $U$-band assuming the cooling break is between the optical and 
X-ray bands at $t > 1000 {\rm ~s}$. The assumed spectral slope in the 
optical is $\beta_O = 0.46$. Although the scatter looks slightly 
less for the $\beta_O = 0.46$ case, the RMS residual for both cases 
is 0.45 mag (see below).

As seen in Figure~\ref{fig4}, there appears to be flaring in the UVOT 
lightcurve between $3000-40\,000 {\rm ~s}$. To determine the statistical 
significance of these potential flares, the shifted UVOT photometry 
was fitted with a single power law ($\chi_{red}^2 = 2.59$ \& $rms_{scatter} 
= 0.45$ mag for $\beta_O = 0.96$; $\chi_{red}^2 = 2.40$ \& $rms_{scatter} 
= 0.45$ mag for $\beta_O = 0.46$). Based on this fit, the 
fluctuations appear real. If the fluctuations represent
flares and a single power law is fit to the bottom of the photometry 
data (i.e. a line is placed through the lightcurve while ignoring the 
three flares), the flaring becomes even more 
significant ($\chi_{red}^2 = 5.16$ \& $rms_{scatter} = 0.50$ mag for 
$\beta_O = 0.96$). A further check was made to determine if 
the fluctuations were caused by instrumental or other effects. Light 
curves for seven comparison stars (which were selected based on 
similar magnitudes to the afterglow's) were constructed in each 
filter after shifting to the $U$-band as described above. For each 
comparison star a weighted mean magnitude and the residual about the 
weighted mean was computed. The RMS residual 
of all the observations of all the stars is 0.09 mag, which is much 
smaller than the RMS residual about the power law fit to the 
afterglow's light curve. We therefore conclude that the comparison 
stars do not show the same fluctuations as the optical afterglow and 
that the fluctuations are intrinsic to the afterglow. Although 
optical flaring has been reported for long bursts 
\citep[cf.][]{HST2002, JP2004}, this is the first short burst with 
optical flaring.

\subsection{Spectral Analysis}
Joint spectral analysis was carried out using the BAT data between 
15 and $150 {\rm ~keV}$ and the KONUS data from 20 to $3000 {\rm ~keV}$. The spectra 
were fit by a power law model with an exponential cut off: 
$dN/dE \sim E^{-\Gamma} exp^{-(2-\Gamma)E/E_p}$ where $E_p$ is the 
peak energy of the $\nu F_{\nu}$ spectrum and $\Gamma$ is the photon 
index. A fit to the Band (GRBM) model was also attempted. 
No statistically 
significant high energy power-law tail was established in any 
fitted spectrum. The time-integrated spectrum is well fit with 
$\Gamma = 0.61\pm0.10$ and $E_p = 947^{+224}_{-173} {\rm ~keV}$ 
($\chi^2 = 115/123 {\rm dof}$). Using these parameters, the calculated 
burst fluence is $1.29^{+0.15}_{-0.31} \times 10^{-5} {\rm ~erg\,cm^{-2}}$ 
and the peak flux is $5.99^{+0.10}_{-1.59} \times 10^{-5} {\rm ~erg\, 
cm^{-2}\,s^{-1}}$, as measured from $T_0 = 0.475 {\rm ~s}$ on a $16 {\rm ~ms}$ 
timescale. All errors are at the 90\% confidence level. Joint fits 
were also made to the same cut-off power law model for three 
specific time intervals denoted on Figure~\ref{fig3}. The results are 
summarized in Table~\ref{tab3}. Significant spectral evolution in both the 
low-energy photon index and $E_p$ is evident. 

\begin{deluxetable}{lrrlrl}
\tablecolumns{6}
\tabletypesize{\small}
\tablecaption{Parameters of a joint BAT-KONUS fit to 
	a model consisting of a power law with an exponential 
	cut off.\tablenotemark{a} 
	\label{tab3}}
\tablewidth{0pt}
\tablehead{
  \colhead{$t\,(s)$} &
  \colhead{$f_{RN}$} &
  \colhead{$\Gamma$} &
  \colhead{$E_{p}$} &
  \colhead{$\chi^{2}/dof$} &
  \colhead{Prob}
}
\startdata
$-0.085\,-\,0.107$ (A) & 1.0 & $-0.01^{+0.13}_{-0.14}$ & $844^{+110}_{-95}$
   & $95.4/89$ & 0.30 \\
    & $0.80^{+0.12}_{-0.11}$ & $ 0.12^{+0.14}_{-0.15}$ & $831^{+112}_{ -97}$
   & $89.1/89$ & 0.45 \\
$\phm{-} 0.107\,-\,0.171$ (B)  & 1.0 & $-0.22^{+0.45}_{-0.61}$ & $237^{ +67}_{ -58}$
   & $73.3/64$ & 0.20 \\
    & $0.96^{+0.36}_{-0.23}$ & $-0.23^{+0.46}_{-0.62}$ & $233^{ +93}_{ -59}$
   & $73.2/63$ & 0.18 \\
$\phm{-} 0.171\,-\,8.363$ (C)  & 1.0 & $ 0.80^{+0.10}_{-0.13}$ & $921^{+474}_{-274}$
   & $122.4/125$ & 0.55 \\
    & $1.18^{+0.16}_{-0.18}$ & $ 0.74^{+0.11}_{-0.14}$ & $990^{+521}_{-286}$
   & $119.8/124$ & 0.59 \\
$-0.085\,-\,8.363$     & 1.0 & $ 0.64^{+0.08}_{-0.10}$ & $924^{+226}_{-165}$
   & $116.0/124$ & 0.68 \\
    & $1.07^{+0.12}_{-0.12}$ & $ 0.61^{+0.09}_{-0.11}$ & $947^{+224}_{-173}$
   & $115.1/123$ & 0.68 \\
\enddata
\tablenotetext{a}{Two sets of fit results are shown for each time interval
  ($t$). In the first row for each time interval, the joint fitting was
  done without renormalization between BAT and KONUS. In the second rows,
  the relative normalization ($f_{RN}$) is allowed to be a free parameter. In
  the case in which $f_{RN}$ is not statistically consistent with unity, the
  difference is attributed to the differences in the low energy parts of the
  BAT and KONUS spectra. The photon index is $\Gamma$, $E_{p}$ is the peak
  energy, $\chi^{2}/dof$ is the goodness of fit, and Prob is the probability.}
\end{deluxetable}

We also constructed a spectral energy distribution (SED) of the 
optical and UV afterglow for $14 {\rm ~ks}$ after the trigger. This time was 
chosen because the lightcurve is relatively well-sampled in the 
different UVOT bands around that point, requiring the minimum of 
interpolation in constructing the SED. For the $B$, $U$, and UVW1 
bands respectively, a power-law was fitted to the data points 
between 4000 and $40\,000 {\rm ~s}$ and the count rates at $14 {\rm ~ks}$ were 
interpolated from the fitted curves. For the $V$, UVM2, and UVW2 
bands, for which there was only one point per filter in the 
$4000-40\,000 {\rm ~s}$ time range, an average power-law 
decay fitted jointly 
to $B$, $U$, and UVW1 was renormalized to the single points in each 
of $V$, UVM2, and UVW2. The count rates at $14 {\rm ~ks}$ in these filters 
were then determined from these extrapolated curves. The count 
rates from all six filters were used to make spectral files for 
fitting with XSPEC using the latest UVOT response matrices 
(version 102). The errors on the count rates consisted of a sum in 
quadrature of the photometric error and a 10\% systematic error 
from the calibration uncertainty on the filter effective areas.

An XRT spectrum for the time interval $4100-24\,000 {\rm ~s}$ was 
extracted for joint fitting with the UVOT, and renormalized to the 
total count rate expected at 14 ks. Fitted by itself, the XRT 
spectrum was slightly softer than that obtained from the whole time 
interval, with a (Galactic absorbed) power-law 
slope of $1.97\pm0.09$. 
The XRT and UVOT spectra were then fitted simultaneously, with a 
model consisting of power-law absorbed by Galactic gas and dust as 
well as dust intrinsic to the GRB host.

\begin{figure}
\includegraphics[angle=-90,scale=0.68]{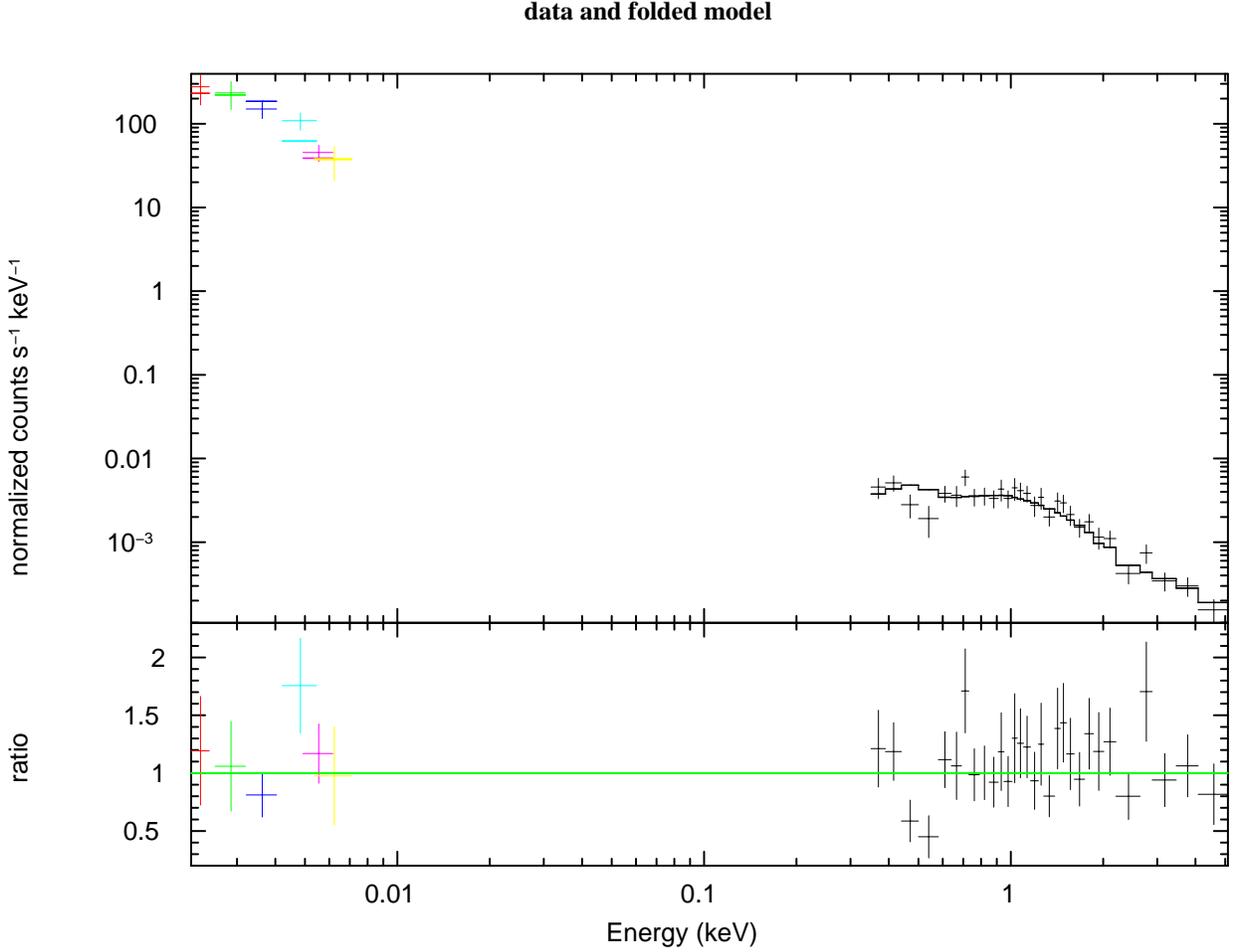}
\caption{Top panel: Combined UVOT and XRT spectral energy distribution 
(SED) of the afterglow at $14 {\rm ~ks}$ after the burst (red = $V$, green = $B$, 
dark blue = $U$, light blue = UVW1, pink = UVM2, yellow = UVW2, black = 
X-ray). The model fit is over plotted; this model, which was fitted to 
both the UVOT and XRT data simultaneously, consists of a single 
power-law with a slope of $2.03\pm0.04$ and Galactic neutral gas absorption 
and dust extinction (the full model parameters are listed in Table~\ref{tab4}). 
This power-law is softer that that fitted to the complete XRT dataset 
(which is biased towards earlier times due to the declining count rate). 
No absorption or extinction intrinsic to the GRB host is required in the 
fit. Bottom panel: Ratio of this model to the UVOT and XRT data.\label{fig5}}
\end{figure}

A good fit was obtained with the power-law with Galactic dust and 
gas alone; no dust intrinsic to the source was required. Since we 
cannot fit the redshift of any dust feature we are unable to use 
the SED fitting to constrain the redshift of the GRB in this case. 
The full parameters of the joint UVOT-XRT fit are given in Table~\ref{tab4}, 
and the UVOT and XRT spectra are plotted with the model and the 
ratio of data to the model in Figure~\ref{fig5}.

\begin{deluxetable}{ll}
\tablecolumns{2}
\tabletypesize{\small}
\tablecaption{Parameters of a power-law fit to the UVOT-XRT spectrum
	with Galactic dust and gas plus dust absorption intrinsic to the
	GRB host.\label{tab4}}
\tablewidth{0pt}
\tablehead{
  \colhead{Component} &
  \colhead{Parameter Values}
}
\startdata
Power-law ($\gamma$) & $2.03 \pm 0.04$ \\
Normalization & $5.5 \pm 0.4 \times 10^{-5}$ ${\rm photons\,cm^{-2}\,
s^{-1}\,keV^{-1}}$ (@ 1 keV) \\
Neutral Galactic gas absorption ($N_{H}$) &${\rm 4.74 \times 10^{20} 
cm^{-2}}$ (fixed) \\
Dust type & Milky Way \\
E(B-V) & 0.0625 mag (fixed) \\
z & 0 (fixed) \\
$\chi_{red}^2$ & 1.142 (34 dof) \\
\enddata
\end{deluxetable}

\section{Discussion}
\subsection{Prompt Gamma-Ray Emission}
\begin{figure}
\plotone{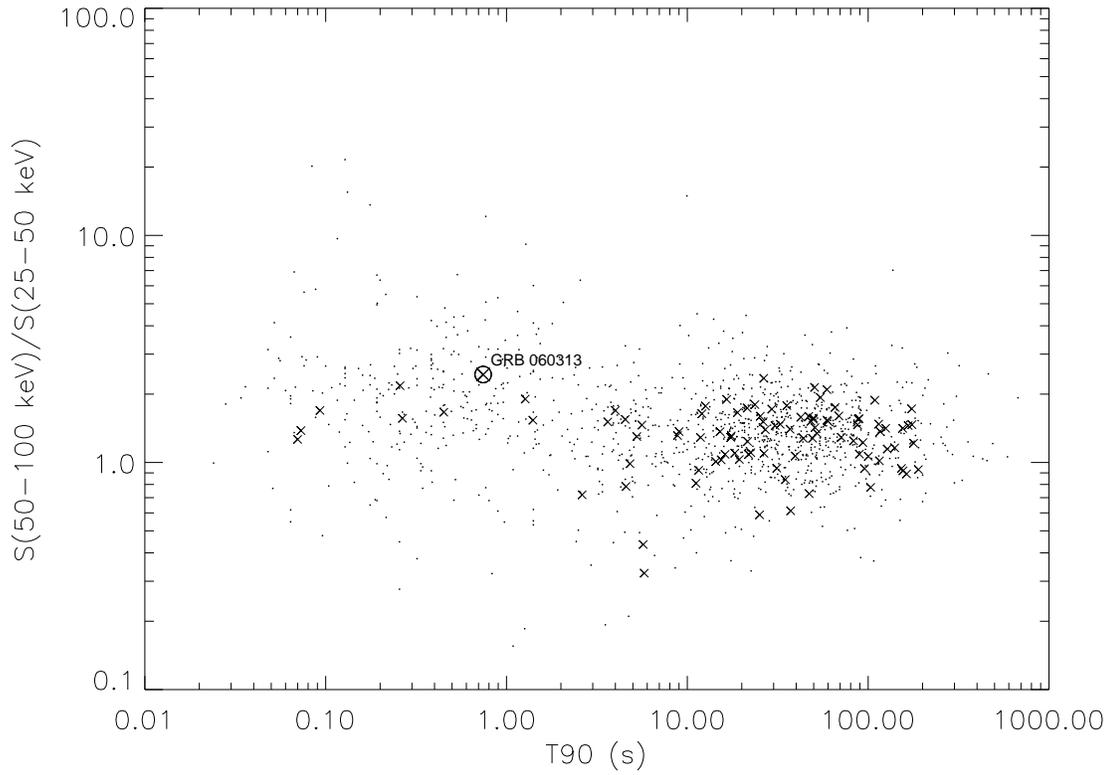} 
\caption{Hardness ratio versus $T_{90}$ for the BAT and BATSE bursts. 
This plot shows how the spectral hardness and duration of GRB 060313 
compares to previous BAT bursts (X symbols), and BATSE bursts (dots). 
One can see that GRB 060313 is the hardest burst yet detected by BAT, 
although it falls well within the BATSE hardness distribution.\label{fig6}}
\end{figure}

Although GRB 060313 is the hardest of the Swift bursts (based on 
the (50-100 keV)/(25-50 keV) ratio), it is just above the average 
hardness for BATSE short bursts (see Figure~\ref{fig6}). Since the BAT 
lightcurve of GRB 060313 is bright and therefore the statistical 
significance is high, we can be confident that many of 
the multi-peaked pulses in the lightcurve are real. Most BATSE short 
bursts, even when as bright as GRB 060313, did not reveal as much 
structure as is evident in this burst. The burst fluence as measured 
in the $15 {\rm ~keV} - 3 {\rm ~MeV}$ band was $1.29^{+0.15}_{-0.31} \times 
10^{-5} {\rm ~erg\,cm^{-2}}$, providing an isotropic gamma-ray energy 
release of $3.4 \times 10^{52} {\rm ~ergs}$ for an assumed burst of 
redshift 1.

The spectrum during the burst had a positive spectral index during 
the first $0.19 {\rm ~s}$ of the burst ($f_\nu \propto \nu^{1.06}$) and the 
time averaged low energy spectrum was $f_\nu \propto 
\nu^{0.29\pm0.07}$. These spectra can be produced in the 
synchrotron-self-Compton model for gamma-ray generation 
\citep{KP2006}. The high degree of variability of the gamma-ray 
lightcurve is probably due to a highly variable outflow from the 
central explosion.

\subsection{Afterglow}
The X-ray lightcurve has a profile similar to the canonical profile 
described in the literature \citep{ZB2006} with the exception that 
segment I (the time immediately after the prompt emission) has a much 
shallower decay slope ($\sim 2\times$ less than the canonical). This 
XRT decay slope of $\alpha = 1.25\pm0.15$ beginning $\sim 100 {\rm ~s}$ after 
the burst trigger is much less than the early decay slopes seen at 
similar times for some short bursts: GRBs 050724 \citep[$\alpha = 7$;][] 
{BS2005b}, 050813 \citep[$\alpha = 2.05\pm0.20$;][]{RJ2006}, 
051210 \citep[$\alpha = 2.57\pm0.11$;][]{LPV2006}, and 051227 
\citep[$\alpha = 2.2\pm0.2$;][]{BL2006}. The decay slope is more 
consistent with the short bursts: GRBs 050509B 
\citep[$\alpha = 1.20^{+0.09}_{-0.08}$;][]{GN2005}, 050709 
\citep[$\alpha \lesssim 1$;][]{FD2005}, and 051221A 
\citep[$\alpha = 1.3\pm0.7$;][]{CM2005}.

The initial XRT lightcurve has flaring in the first $1000 {\rm ~s}$ after the 
trigger. The spectrum during this period is harder (photon index of 
$1.53\pm0.10$) than the best spectral fit of the late afterglow (photon index 
of $1.96\pm0.09$). Flares in the early lightcurve of the short GRBs 050724 
\citep{BS2005b, CS2006, GD2006}, 051210 
\citep{LPV2006}, and 060121 \citep{LA2006} have also been seen. 
Flaring in GRBs has been attributed to long lasting activity by the 
central engine with internal shocks continuing for hundreds of 
seconds \citep{BDN2005b, FA2006a, FA2006b, PC2006c, ZB2006, DPM2006}. 
Although the 
decay slope of GRB 060313 is consistent with GRBs 050509B, 050709, 
and 051221A, no early 
flaring was seen in these three bursts (Fox {\etal} (2005) report the 
possibility of a flare at early times for GRB 050709, however, the 
fitting of the lightcurve at early times was difficult since only one 
data point was available in the first day after the burst). Between 
$1000-3000 {\rm ~s}$ no data was obtained due to occultation by the Earth; 
therefore, it is unknown what the lightcurve profile is at this time. 
However, this region in the lightcurve is consistent with either 
segment II or V in Zhang {\etal} (2006), therefore, there is clear 
evidence that there is some kind of energy injection

The optical and UV lightcurve is the earliest ($< 80 {\rm ~s}$) optical/UV 
detection of a short GRB to date. The early UVOT decay slope, 
beginning $\sim 100 {\rm ~s}$ after the burst trigger, is 
$\alpha = 0.13\pm0.28$ 
which is much flatter than the XRT decay slope over the same period. 
In addition, unlike the XRT, the UVOT lightcurve contains no flaring 
during this period. At $t > 3000 {\rm ~s}$, the UVOT lightcurve is steeper 
($\alpha = 0.43\pm0.13$) and includes three flares superimposed
on the overall lightcurve 
while the XRT lightcurve has no flaring. No other short burst has so far been 
observed to manifest optical flaring at such early times.

The XRT spectrum during the first $1000 {\rm ~s}$ of the burst was 
proportional to $\nu^{-0.53\pm0.10}$ in the $0.3-10 {\rm ~keV}$ energy 
band. The spectrum in this entire energy band was a single power-law 
meaning it did not contain any characteristic synchrotron frequency. 
During this same period, the XRT lightcurve fell off as $t^{-1.25}$ 
signifying that the synchrotron peak frequency was below the XRT band 
or less than $0.3 {\rm ~keV}$. This indicates that the synchrotron 
cooling frequency was either less than $0.3 {\rm ~keV}$ or above $10 
{\rm ~keV}$. If the cooling frequency were to be below $0.3 {\rm 
~keV}$ then the XRT spectrum would be $\nu^{-p/2}$ (where $p$ is the 
electron power-law distribution index) giving a value for $p = 
1.06\pm0.2$ which is highly unlikely (it is much too small for the 
X-ray spectrum of GRB afterglows). We therefore conclude that the 
cooling frequency for $t < 1000 {\rm ~s}$ was above $10 {\rm ~keV}$ 
\citep[cf.][]{SPN1998, CL1999, RE2005}.
An X-ray lightcurve due to shocks in a uniform density 
circum-stellar medium (CSM) is expected to fall off as $t^{-0.8}$, for 
a photon index of 1.53, instead of the observed $t^{-1.25}$ decay. The 
observed lightcurve decay is consistent with a wind-like density 
stratification for the CSM. However, the subsequent decline of the 
X-ray lightcurve and the spectrum are seriously at odds with this 
possibility and therefore we do not consider it to be a viable 
solution. The steeper than expected fall off of the lightcurve could be 
due to some combination of a structured jet, radiative loss of energy, 
and decreasing microphysics parameters \citep[$\epsilon_e$ \& $\epsilon_B$; 
cf.][]{PA2006b}.

The flux in the $0.3-10 {\rm ~keV}$ band at $100 {\rm ~s}$ from the forward shock, 
assuming the XRT band below the synchrotron cooling frequency and a burst 
redshift of 1 which provides a luminosity distance ($D_{28}$) $\sim1$,
is (eq. B7 from Panaitescu \& Kumar (2000) is used since we assume
$\nu_i < \nu < \nu_c$ and a homogeneous external medium) 
\begin{equation}
f_x \approx 5\times10^{-9}\,E_{52}^{5/4}\,n^{1/2}\,\epsilon_{e,-1}\,
\epsilon_{B,-2}^{3/4}\quad {\rm erg\,cm^{-2}\,s^{-1}},
\end{equation}
where $E_{52}$ is the isotropic equivalent energy in units of $10^{52} 
{\rm ~ergs}$, $\epsilon_{e,-1} = \epsilon_e /0.1$ ($\epsilon_e(p-1)/(p-2)$ 
is the energy fraction in electrons), 
$\epsilon_{B,-2}$ is the energy fraction in the magnetic field, $n$ is 
the density of the CSM, and we used $p = 2.06$ as suggested by the 
early X-ray spectrum\footnote{Since $f_\nu \propto \nu^{-0.53}$, then 
$p=2.06\pm0.2$. The lower bound (1.86) would lead to a negative energy 
fraction of electrons which is an unphysical event.}, in calculating the 
numerical values in the above equation. Note that the error in the 
burst redshift has a very small effect on the X-ray flux formula given 
above as $f_x$ has almost a linear dependence on the isotropic burst 
energy and therefore the effect of uncertainty in distance on $E_{52}$ 
and $f_x$ nearly cancel each other. Substituting the observed value of 
the X-ray flux of $3 \times 10^{-10} {\rm ~erg\,cm^{-2}\,s^{-1}}$ in the above 
formula we find
\begin{equation}
E_{52}^{5/4}\,n^{1/2}\,\epsilon_{e,-1}\,\epsilon_{B,-2}^{3/4}\approx 6 
\times 10^{-2}.
\end{equation}
Moreover, applying the constraint that the cooling frequency is greater 
than $10 {\rm ~keV}$ at $10^3 {\rm ~s}$ (because the photon spectral index is 1.53) we 
obtain 
\begin{equation}
E_{52}^{1/2}\,n\,\epsilon_{B,-2}^{3/2} < 5 \times 10^{-3}.
\end{equation}
Equation (3) follows from the expression of cooling frequency given in the 
literature (eq. 27 from Panaitescu \& Kumar (2000) assuming $Y \ll 1$).
Combining equations (2) and (3) we find: $\epsilon_{e,-1}\,E_{52} > 
1$. Moreover, for $p = 2.06$ the fraction of energy of the shocked fluid 
in electrons is equal to $1.76\epsilon_{e,-1}$ which cannot be greater 
than 1, and therefore $\epsilon_{e,-1} < 0.6$ and $E_{52} > 1.8$. The 
$15 {\rm ~keV}-3 {\rm ~MeV}$ $\gamma$-ray fluence suggests that the isotropic equivalent 
energy in photons, assuming $z = 1$, was $E_{52} = 3.4$, which is 
consistent with the above derived constraint on the energy in the 
explosion of $E_{52} > 1.8$. Since the radiative efficiency for the 
prompt $\gamma$-ray emission must be less than 100\%, and likely of order 
a few tens of percent, this suggests $E_{52} \sim 6$, and $\epsilon_{e,-1} 
\sim 0.5$. Substituting these values in equation (2) we find
\begin{equation}
n^{1/2}\,\epsilon_{B,-2}^{3/4} \approx 1 \times 10^{-2}.
\end{equation}
We next make use of constraints on the synchrotron injection frequency (the 
synchrotron frequency at which the bulk of the electrons which are injected 
by the shock front radiate)
which is given by \citep[e.g. eq. 22 in][]{PK2000}
\begin{equation}
\nu_i = 306\,E_{52}^{1/2}\,\epsilon_{e,-1}^2\,\epsilon_{B,-2}^{1/2}\, 
t_2^{-3/2} \quad {\rm eV}.
\end{equation}
Since the XRT lightcurve is falling off as $t^{-1.25}$ starting from 
the very first XRT data point at $t_2 = 1$ (where $t_2 = t/100 {\rm ~s}$), we 
conclude that $\nu_i < 0.3 {\rm ~keV}$ at this time. If $\nu_i > 0.3 {\rm ~keV}$,
then the XRT lightcurve would be rising with time as $t^{1/2}$ 
\citep[cf.][]{PK2000}. Using equation (5) for 
$\nu_i$ we therefore conclude that $\epsilon_{B,-2} < 2.6$; $E_{52} = 
6$ \& $\epsilon_{e,-1} = 0.5$ were used in this estimation. Moreover, at 
$t = 1 {\rm ~ks}$ or $t_2 = 10$, $\nu_i$ should be greater than 
$2 {\rm ~eV}$ since the 
UVOT lightcurve is flat during the time interval $100-1000 {\rm ~s}$. This 
constraint, and equation (5), leads to $\epsilon_{B,-2} > 0.1$.  From 
these upper and lower limits on $\epsilon_{B,-2}$ it seems reasonable 
to infer that $\epsilon_{B,-2} \sim 1$, which when substituted into 
equation (4) gives the density of the CSM to be $n \sim 10^{-4} 
{\rm ~cm^{-3}}$. This is consistent with the combined UVOT/XRT SED which 
reveals that no intrinsic dust is required to model the environment of 
the burst. This appears that the burst went off in a low density medium 
that is usually found in bubbles or the outskirts of galaxies.

If the UVOT band lies below $\nu_i$, the lightcurve is expected to rise 
as $t^{1/2}$ \citep[cf.][]{PT1999,PK2000}. However, the lightcurve seems 
to be nearly flat for the 
initial 10 minutes or so, thus, we conclude that $\nu_i$ is below the 
UVOT band. It is likely that the same process that caused a steeper 
fall off of the early X-ray lightcurve, by $\sim t^{0.45}$, could have 
also flattened the rise of the early UVOT lightcurve from the expected 
$t^{1/2}$ to $t^{\sim 0}$. The general decline of the optical lightcurve 
between 10 and $40 {\rm ~ks}$ is roughly the same as the early X-ray lightcurve 
and that is consistent with the forward external shock model.

The X-ray lightcurve for $t > 4 {\rm ~ks}$ is steeper than the early 
lightcurve by $t^{1/4}$ and that is almost certainly a result of the 
cooling frequency dropping below the XRT band; the spectrum during 
this stage also became softer as expected for a cooling transition. 
However, the flux at $t > 4 {\rm ~ks}$ is larger by a factor $\sim 4$ compared 
with the extrapolation of the earlier X-ray lightcurve. This suggests 
quite a substantial amount of energy being added to the forward shock, 
almost twice the amount of the initial energy in the explosion, during 
the time period of 1 to $4 {\rm ~ks}$. It is also possible that the enhanced 
X-ray flux could be due to the GRB ejecta running into a denser shell 
of CSM.

There are fluctuations in the UVOT lightcurve between 4 and $40 {\rm ~ks}$ 
whereas during this period the X-ray lightcurve has a smooth powerlaw 
decline. This interesting behavior is not expected for energy being 
added to the decelerating forward shock (due to late time central 
engine activity, for instance) whereas this is consistent with 
fluctuations in the density of the CSM. The observed flux at 
frequencies above the cooling frequnecy ($\nu_c$) is very insensitive 
to the CSM density structure and this could explain the smoothly 
declining XRT lightcurve. The flux at frequencies between $\nu_i$ and 
$\nu_c$ varies with density as $n^{1/2}$ \citep[cf.][]{PT1999,PK2000}, 
and hence density 
fluctuations will be reflected in the optical lightcurve if 
$\nu_{opt} < \nu_c$; the amplitude of the UVOT flux variation 
corresponds to a factor of about 2 variations in CSM density, and the 
short timescale for the variability suggests small scale fluctuation 
in the CSM. We note that the early X-ray lightcurve ($t < 10^3 {\rm ~s}$), 
when the XRT band was below $\nu_c$, also showed fluctuations, 
lending additional credence to the interpretation of CSM density 
fluctuation.

The variations in the XRT and UVOT lightcurve occur on very short 
timescales ($dt/t \ll 1$, where $dt$ is the variability timescale). 
These small amplitude fluctuations can 
arise in the external shock with $dt/t < 1$; for instance, when 
the shock front crosses a density clump of size much less than 
$r/\Gamma$ ($r$ is the radius of the external shock and $\Gamma$ 
is the Lorentz factor) it will increase the observed flux a bit. 
This increase will last for a time $dt < t$ where $dt/t$ is 
determined by the clump size, and the amplitude of fluctuation by 
its mass.

This short burst offers conflicting evidence as to the nature 
of the progenitor and its surrounding medium. On the one hand the 
afterglow data suggests that the burst went off in a low density 
medium and on the other hand fluctuations in the UVOT lightcurve 
and a lack of corresponding fluctuation in the X-ray data suggests 
that the CSM density varied by a factor of order 2 on a length scale 
of a few times $10^{17} {\rm ~cm}$.

The low density of the circum-stellar medium has fluctuations that 
one normally associates with the surroundings of a massive star and 
inhomogeneities of a windy medium. However, the prospect of the 
progenitor being a massive star is unlikely. The mean density of the 
medium for this burst is small ($\sim 10^{-4} {\rm ~particles\,cm^{-3}}$) for 
a massive star. In addition, although it can not be ruled out, no host 
galaxy was found at the location of GRB 060313. Instead, a search in 
the extragalactic databases indicates that there are 12 galaxies 
within 5\arcmin ~of GRB 060313, of which 6 are within 2\arcmin ~and one 
within 30\arcsec ~(B042408.28-105721.7). Could this possibly be a 
galaxy cluster with which the GRB is associated? Currently there are 
no known redshifts or color information on these galaxies; therefore, 
it can not be determined for certain. However, if we assume that there 
is no host galaxy at the position of the GRB and that it is part of this 
potential galaxy cluster, it is unlikely that a massive star can travel 
such great distances. Massive stars live for about 10 million years, 
and their peculiar velocity is not more than a few tens of ${\rm km\,s^{-1}}$. 
Therefore, in their lifetime, a massive star can only travel a distance 
of less than a few hundred pc.
 
In contrast, the natal kick of a neutron star ($\sim 200-1000 {\rm ~km\,s^{-1}}$) 
and the typical life times of compact binaries ($\sim 0.1-1 {\rm ~Gyr}$) 
allow for DNS and NS-BH binaries to travel great distances. Assuming a 
redshift for GRB 060313 of $z = 0.16$ \citep[for GRB 050709 - the closest 
measured redshift for a short GRB;][]{FD2005}, $z = 0.35$ \citep[the 
average measured redshift for short GRBs;][]{LPV2006}, $z = 0.55$ 
\citep[for GRB 051221A - the farthest measured redshift for a short GRB;][]{SA2006}, 
and $z = 1.1$ (for the UVOT upper redshift limit), a 
compact binary progenitor could have been associated with a galaxy that 
is $\lesssim 6.2\arcmin, 3.5\arcmin, 2.7\arcmin$, and $2.1\arcmin$ ~away, 
respectively. These distances are consistent with the distances to some 
of the galaxies in the potential cluster. Without the early lightcurves, 
particularly the UV/optical lightcurve, we would have automatically 
assumed that the progenitor of this GRB was a compact binary merger. 
However, the data suggests that GRB 060313 may arise from a mechanism 
other than the established compact binary model.

An alternative possibility is that the rapid variabilities in the UVOT 
lightcurve are produced by late-time central engine activity similar to 
those produced in late-time X-ray flares seen in many other GRBs 
\citep[e.g.][]{BDN2005b, ZB2006, LA2006}. Such a possibility 
has been raised to discuss other rapid optical variabilities of pre-Swift 
afterglows with $\Delta{t} < t$ \citep[e.g.][]{IKZ2005}. In this scenario, 
no CBM density fluctuation is required. The lack of variability in the 
X-ray band may be due to the fact that the $E_p$'s of these late flares 
are very low (i.e. in the UV/optical regime), so that their contributions 
to the X-ray band are negligible compared with the afterglow level. 
Within the late internal shock scenario, soft flares are in principle 
possible given the combination of a low luminosity and a large Lorentz 
factor as a result of less baryon loading at later times \citep{ZM2002}. 
The challenge is how to restart the central engine at very late epochs 
in short GRBs. The same problem has also been encountered when 
interpreting the multiple late time flares of GRB 050724 
\citep{BS2005b}, and some suggestions have been made to explain its
behavior \citep[e.g.][]{BS2005b, PAZ2006, PZ2006, DZG2006, FJ2006}.

\section{Conclusion}
The multi-wavelength lightcurve of the SHB GRB 060313 reveals a 
multi-peaked structure in the $\gamma$-ray, hard X-ray, X-ray, and 
UV/Optical bands at different epochs of the prompt and afterglow 
emissions. The hard X-ray spectrum during the prompt emission had a 
positive spectral index consistent with the synchrotron-self-Compton 
model for $\gamma$-ray generation. The large number of peaks in the 
$\gamma$-ray lightcurve is possibly the result of the central engine 
generating an extremely variable outflow. 

The early ($70-1000 {\rm ~s}$) X-ray temporal decay slope ($\alpha = 1.25\pm0.15$) 
is exceptionally steep for the observed X-ray spectrum and manifests 
flaring components superimposed on the lightcurve. During the same period 
the UV/Optical lightcurve reveals a flat decay profile 
($\alpha = 0.13\pm0.28$) with no flaring. This behavior can be attributed 
to a combination of a structured jet, radiative loss of energy, and 
decreasing microphysics parameters. The X-ray spectrum indicates that the 
cooling frequency was not much greater than $10 {\rm ~keV}$ at $1000 {\rm ~s}$ after the 
burst. At later times ($> 3000 {\rm ~s}$) the UV/Optical lightcurve manifests 
flaring on top of a decaying profile ($\alpha = 0.43\pm0.13$) while the 
X-ray exhibits no evidence of flaring on top of a lightcurve that is 
steeper than at earlier times. An interpretation of the data points to a 
burst that occurred in a low density medium and that the CSM density 
varied by a factor of order 2 on a length scale of a few times $10^{17} 
{\rm ~cm}$. Although the collapse of a massive star as the progenitor mechanism 
is doubtful (though it can not be ruled out), the favored model of a 
compact binary merger also creates problems in explaining the burst's 
behavior. One explanation for this behavior is that at least some short bursts 
may be the result of a different mechanism other than the traditional compact 
binary model. An alternate explanation is that late-time central engine 
activity is injecting energy into the UV/optical regime but not into the X-ray
regime.

This is the first short burst that has manifested this kind of behavior. 
Because the sample size is small, more observations of these objects are 
needed in order to determine whether GRB 060313 is the norm or anomalous 
for short GRBs. In addition, more detailed theoretical modeling is 
needed to establish a clearer picture of the mechanism.

\acknowledgments
This work is sponsored at Penn State by NASA contract NAS5-00136, at 
University College London, Mullard Space Science Lab and University of 
Leicester by funding from PPARC, and at Osservatorio Astronomico di 
Brera by funding from ASI on grant number I/R/039/04. The KONUS-Wind 
experiment is supported by a Russian Space Agency contract and RFBR 
grant 06-02-16070. We appreciate the thorough review of this paper by
an anonymous referee. We gratefully appreciate the contributions of all 
members of the Swift team. PR would like to acknowledge Heidi Roehrs 
for her contributions.

\clearpage

\end{document}